\documentclass[prl,aps,twocolumn,superscriptaddress]{revtex4-1}
\usepackage{latexsym}
\usepackage{amssymb}
\usepackage{amsmath}
\usepackage{amsbsy}
\usepackage{mathtools}
\usepackage{mathrsfs}
\usepackage{physics}
\usepackage[pdftex]{graphicx}
\usepackage{epstopdf}
\usepackage[usenames, dvipsnames]{color}
\usepackage{soul}
\usepackage{xcolor}
\usepackage{graphicx}
\usepackage[normalem]{ulem}
\usepackage{dcolumn}
\usepackage{bm}
\usepackage[english]{babel}
\usepackage{notes2bib}
\usepackage{etoolbox}

\newcommand{\uami}{Departamento de F\'isica,
Universidad Aut\'onoma Metropolitana
Iztapalapa, Av. San Rafael Atlixco 186, Col. 
Vicentina, 09340 Ciudad de M\'exico, M\'exico}

\newcommand{\uama}{Departamento de Ciencias B\'asicas,
Universidad Aut\'onoma Metropolitana
Azcapotzalco, Av. San Pablo 180, Col. 
Reynosa Tamaulipas,  Ciudad de M\'exico, M\'exico}

\newcommand{\unam}{Departamento de Sistemas Complejos, Instituto de F\'isica, Universidad Nacional Aut\'onoma de M\'exico, Apartado Postal 20-364 01000 Ciudad de M\'{e}xico, M\'{e}xico}
\begin{document}

\title{Method for finding the exact effective Hamiltonian of
time driven quantum systems}

\author{J. C. Sandoval-Santana}
\author{V. G. Ibarra-Sierra}
\affiliation{\uami}
\author{J.L. Cardoso}
\author{A. Kunold}
\affiliation{\uama}
\author{P. Roman-Taboada}
\author{G. G. Naumis}
\affiliation{\unam}


\begin{abstract}
Time-driven quantum systems are important in many different fields of physics like cold atoms, solid state, optics, etc. Many of their properties are encoded in the time evolution operator which is calculated by using a time-ordered product of actions. The solution to this problem is equivalent to find an effective Hamiltonian. This task is usually very complex and either requires approximations,  or in very particular and rare cases, a system-dependent method can be found. Here we provide a general scheme that allows to find such effective Hamiltonian. The method is based in using the structure of the associated Lie group and a decomposition of the evolution on each group generator.  The time evolution is thus always transformed in a system of ordinary non-linear differential equations for a set of coefficients. In many cases this system can be solved by symbolic computational algorithms. As an example, an exact solution to three well known problems is provided. For two of them, the modulated optical lattice and Kapitza pendulum, the exact solutions, which were already known,  are reproduced. For the other example, the Paul trap, no exact solutions were known. Here we find such exact solution, and as expected, contain the approximate solutions found by other authors. 
\end{abstract}

\maketitle

During the last years there has been an ever increasing interest in studying time-driven quantum systems \cite{PhysRevX.4.031027} (TDQS). Among the reasons for this spark of interest, one can mention the possibility of tailoring time driven potentials using cold-atoms \cite{Carleo2017} or optically irradiated  2D materials \cite{Lopez2008,Foa2014}, as well as for quantum entanglement problems \cite{Nahum2017}. Furthermore, it has been found that new and interesting topological properties arise for periodic driven systems \cite{Roman2017}. As a matter of fact, these properties can also be found in 2D materials, as is the case of graphene \cite{Low2012,naumis-review}. Also, quantum-quenching has become a mainstream subject of research \cite{Guardado2018}.  In almost all of these kind of systems \cite{PhysRevX.4.031027}, the Hamiltonian $H(t)=H_0+V(t)$ is written as a time-independent Hamiltonian ($H_0$) plus a time-dependent potential ($V(t)$). Among the most important cases, is the one of a periodic $V(t)$. Here
we will consider such case, with $V(t)$ having a period $T$.  

The TDQS properties are thus calculated by using the time evolution operator  $U(t)=\mathcal{T}\mathrm{e}^{-i\int_0^t dt H(t)/\hbar}$, where $\mathcal{T}$ is the time ordering operator. In the case of periodic potentials, using Floquet theory, one can show that the solution is equivalent to find an effective Hamiltonian $H_\mathrm{e}$ such that \cite{PhysRevX.4.031027},
\begin{equation}\label{Eq:Ueffective}
U(T)=\mathrm{e}^{-i H_{\text{e}} T/\hbar}.
\end{equation}
This effective Hamiltonian encodes all the dynamical information of the system, yet its calculation is not a trivial task. In fact, many few cases allow a closed analytic solution  \cite{PhysRevX.4.031027}. The reason of such difficulty is that usually, $H_0$ and $V(t)$ do not commute. Here we present a general method based on the use of Lie algebras that allows to compute $H_{\mathrm{e}}$.
A great variety of physically relevant Hamiltonians may be addressed by the method proposed here.
As examples we can cite:
the Modulated optical lattice \cite{PhysRevB.34.3625,PhysRevLett.99.220403},
Fastly driven tight-binding chains \cite{ITIN2014822,PhysRevB.96.144301},
Paul trap \cite{refId0},
Quantum wires \cite{PhysRevLett.106.220402},
Graphene \cite{1367-2630-17-9-093039},
Hubbard Hamiltonian \cite{PhysRevLett.95.260404,PhysRevLett.111.175301,
RevModPhys.89.011004}.
Furthermore, Fock space operators have the same algebra than
single particle Hamiltonians \cite{PhysRevX.4.031027}. Therefore, if
the single particle Hamiltonian forms a Lie
algebra so does the second quantization version.
Therefore, the second quantization counterpart of any single particle
Hamiltonian can be addressed in the same way. The method can also be used to find a gauge transformation
so that the Hamiltonian is time-independent \cite{PhysRevA.68.013820,PhysRevX.4.031027}.

A Hamiltonian is said to have
a dynamical algebra
if it can be expressed as the superposition of the 
elements of a finite Lie algebra
$\mathcal{L}_n=\left\{h_1,h_2,\dots,h_n\right\}$ as
\begin{equation}
H=\boldsymbol{a}^\top \boldsymbol{h},\label{ham:01}
\end{equation}
where $\boldsymbol{h}=(h_1,h_2,\dots,h_n)$
and the coefficients
$\boldsymbol{a}^\top=(a_1,a_2,\dots,a_n)$ are in general
time-dependent.
In order for $\mathcal{L}_n$ to be a Lie algebra,
any pair of its elements must meet the following
commutator relation
\begin{equation}
\left[h_i,h_j\right]=i\hbar \sum_{k=1}^n c_{i,j,k}h_k\,\, ,
\end{equation}
where the  structure constants $c_{i,j,k}$
carry all the information regarding $\mathcal{L}_n$.
Part of this information concerns how
the unitary group
generated by $\mathcal{L}$ transforms
any $h_k\in \mathcal{L}_n$.
Indeed, it can be shown that
these transformations depend entirely
on the structure constants.
The elements of the unitary group
$U_k=\exp\left(i\alpha_k h_k/\hbar\right)$
transform $\boldsymbol{h}$ according to
\begin{equation}
U_k\boldsymbol{h}U_k^\dagger=M_k\boldsymbol{h}.
\label{trans:01}
\end{equation}
The matrices $M_k$ can be calculated
by taking the derivative of the left-hand side
of (\ref{trans:01}) with respect to the
parameter
\begin{equation}
\partial_{\alpha_k}
U_k\boldsymbol{h}U_k^\dagger
=\frac{i}{\hbar}
U_k\left[h_k,\boldsymbol{h}\right]U_k^\dagger
=-Q_kU_k\boldsymbol{h}U_k^\dagger,\label{trans:02}
\end{equation}
where  the matrix elements of $Q_k$ are
related to the structure constants by $(Q_k)_{i,j}=c_{i,j,k}$.
By using the condition
$U_k\boldsymbol{h}U_k^\dagger=\boldsymbol{h}$ for
$\alpha_k=0$,
the formal solution to the differential equation
(\ref{trans:02}) is given by
\begin{equation}
U_k\boldsymbol{h}U_k^\dagger
=\exp\left(-Q_k\alpha_k\right)\boldsymbol{h},
\end{equation}
and therefore, the explicit form of the
transformation matrices in Eq. (\ref{trans:01})
is given by
\begin{equation}
M_k=\exp\left(-Q_k\alpha_k\right).
\label{trans:04}
\end{equation}

The time evolution operator $p_t=i\hbar \partial/\partial t$
is transformed as
\begin{equation}
U_kp_tU_k^\dagger=p_t+U_k\left[p_t,U_k^\dagger\right]
=p_t+\boldsymbol{\alpha}^\top I_k\boldsymbol{h},
\label{trans:03}
\end{equation}
where $(I_k)_{i,j} = \delta_{i,j}\delta_{k,j}$. The general form of the evolution operator $U(t)$  for a 
Hamiltonian with a dynamical algebra, can be expressed in terms of either
of the following two forms 
\begin{eqnarray}
\mathcal{U}_A(\boldsymbol{\alpha}) &=&\prod_{k=n}^1 U_k
=\prod_{k=n}^1 \exp\left(i\alpha_k h_k/\hbar\right),
\label{ua:01}\\
\mathcal{U}_B(\boldsymbol{\beta})
&=&\exp\left(\frac{i}{\hbar}\sum_{k=1}^n\beta_kh_k\right)
=\exp\left(
\frac{i}{\hbar}\boldsymbol{\beta}^\top\boldsymbol{h}\right) ,
\label{ub:01}
\end{eqnarray}
where
$U(t)=\mathcal{U}_A^\dagger=\mathcal{U}_B^\dagger$,
$\boldsymbol{\alpha}^\top=(\alpha_1,\alpha_2,\dots,\alpha_n)$
and
$\boldsymbol{\beta}^\top=(\beta_1,\beta_2,\dots,\beta_n)$
are in general time-dependent parameters yet to be determined.
We readily notice that the evolution operator in
(\ref{ub:01}) has the form of (\ref{Eq:Ueffective})
and therefore it follows that
\begin{equation}
\boldsymbol{\beta}^\top(T)\boldsymbol{h}/T=H_{\mathrm{e}}.
\label{he:02}
\end{equation}
Even though in principle it would seem that
a direct path to obtain $H_{\mathrm{e}}$
is to workout
the $\boldsymbol{\beta}(t)$ coefficients, the differential
equations that arise from the evolution operator in
(\ref{ub:01}) are extremely complicated.
Fortunately, the differential equations
ensued from $U_A$ are simpler and render
the $\boldsymbol{\alpha}(t)$ parameters instead.
This, nevertheless, requires
that a relation between the $\boldsymbol{\alpha}(t)$
and $\boldsymbol{\beta}(t)$ parameters be established.

We thus start by determining
the $\boldsymbol{\alpha}(t)$ parameters.
After successively applying the $n$ transformations
in (\ref{ua:01}) to the Floquet operator $H-p_t$
\cite{doi:10.1063/1.4947296}
and using (\ref{trans:01}) and (\ref{trans:03}),
$
\mathcal{U}_A(H-p_t)
\mathcal{U}_A^\dagger=
\mathcal{U}_A(\boldsymbol{a}^\top\boldsymbol{h}-p_t)
\mathcal{U}_A^\dagger=\boldsymbol{u}^\top\boldsymbol{h}-p_t,
$
where
\begin{eqnarray}
\boldsymbol{u}^\top &=& \boldsymbol{a}^\top M_1M_2\dots M_n
-\dot{\boldsymbol{\alpha}}^\top(t)\nu,\label{u:01}\\
\boldsymbol{\nu}^\top &=& I_1M_2\dots M_n+I_2M_3
\dots M_n +\dots+I_n .\label{nu:01}
\end{eqnarray}
In order for $\mathcal{U}^\dagger_A$ to be the
evolution operator, the condition $\boldsymbol{u}=0$
must be fulfilled \cite{doi:10.1063/1.4947296}.
This condition translates into a system of ordinary
differential equations (ODE) for the $\boldsymbol{\alpha}(t)$
parameters that one could in principle attempt to solve.
However, specially for algebras with large dimension,
these equations might be very complex. Therefore,
instead, we solve the simpler system of differential
equations
\begin{equation}
\boldsymbol{\mathcal{E}}=\boldsymbol{\nu}^{-1}\boldsymbol{u}
=\boldsymbol{\nu}^{-1}M_n^\top\dots M_2^\top M_1^\top
\boldsymbol{a}-\dot{\boldsymbol{\alpha}}
=0.\label{ode:01}
\end{equation}
To insure that $\mathcal{U}_A[\boldsymbol{\alpha}(0)]=1$,
the initial
condition $\boldsymbol{\alpha}(0)=0$ must be applied.
Determining $\boldsymbol{\alpha}(t)$
allows us to fully express the evolution
operator in the form (\ref{ua:01}).
In order to find the effective Hamiltonian,
the so obtained evolution operator must
be put in the form of $\mathcal{U}_B$.
Finding the relation between
$\boldsymbol{\alpha}(t)$ and $\boldsymbol{\beta}(t)$
is then essential to working out the effective Hamiltonian.
To obtain such a relation we start by assuming
that both forms of the evolution operator, (\ref{ua:01})
and (\ref{ub:01}), coincide.
This equality should be preserved if we
introduce a dependence in an auxiliary
parameter $\lambda$ by making
$
\mathcal{U}_A[ \boldsymbol{\alpha}(\lambda,t)]
=\mathcal{U}_B(\lambda \boldsymbol{\beta}(t)).
$
It is important to stress that at this point
$\boldsymbol{\alpha}(\lambda,t)$ is both a function of the
parameter $\lambda$ and time.
Conversely, $\boldsymbol{\beta}(t)$ is strictly
a function of time.
When $\lambda=0$, $\boldsymbol{\alpha}(0,t)=0$
since
$\mathcal{U}_B(0)=\mathcal{U}_A[\boldsymbol{\alpha}(0,t)]=1$.
Furthermore, for $\lambda=1$ we recover the original
parameters $\boldsymbol{\alpha}(1,t)=\boldsymbol{\alpha}(t)$.
Taking the derivative
with respect to $\lambda$
of both sides of the previous equation
we get
\begin{multline}
\partial_{\lambda}\, \mathcal{U}_A[\boldsymbol{\alpha}(\lambda,t)]
=  [\partial_{\lambda}  \boldsymbol{\alpha}^\top(\lambda,t)] \nu^\top
\boldsymbol{h} \\
=\boldsymbol{\beta}^\top(t)\boldsymbol{h}
=\partial_{\lambda} \mathcal{U}_B(\lambda\boldsymbol{\beta}(t)), \label{rel:05}
\end{multline}
where $\nu\equiv \nu[\boldsymbol{\alpha}(\lambda,t)]$.
Factorizing $\boldsymbol{h}$,
transposing and inverting $\nu$,
Eq. (\ref{rel:05}) can be recast
in the form of the ODE system of differential
equations for $\boldsymbol{\alpha}(\lambda,t)$
\begin{equation}
\partial_{\lambda} \boldsymbol{\alpha}(\lambda,t) 
=\nu^{-1}[\boldsymbol{\alpha}(\lambda,t)]\boldsymbol{\beta}(t).
\label{rel:01}
\end{equation}
The key element to deduce the relation
between $\boldsymbol{\alpha}(t)$ and $\boldsymbol{\beta}(t)$
is solving this ODE system.
Its solution renders $\boldsymbol{\alpha}(\lambda,t)$
in the form of a function of $\lambda$ and
$\boldsymbol{\beta}(t)$
\begin{equation}
\boldsymbol{\alpha}(\lambda,t)=
\boldsymbol{\alpha}(\lambda,\boldsymbol{\beta}(t)).
\label{rel:02}
\end{equation}
The inverse of (\ref{rel:02}) evaluated in $\lambda=1$
yields the desired relation of
$\boldsymbol{\beta}(t)$
as a function of $\boldsymbol{\alpha}(1,t)$
\begin{equation}
\boldsymbol{\beta}(t)=
\boldsymbol{\beta}[\boldsymbol{\alpha}(1,t)]
=\boldsymbol{\beta}[\boldsymbol{\alpha}(t)].
\label{rel:03}
\end{equation}
Nonetheless, the analytical solution of the ODE system (\ref{rel:01})
or the inverse relation (\ref{rel:03})
might be challenging to work out.
To overcome this difficulty we observe that
$\boldsymbol{\beta}^\top (t)\boldsymbol{h}
=\mathcal{U}_B\boldsymbol{\beta}^\top(t)\boldsymbol{h}\mathcal{U}_B^\dagger
=\boldsymbol{\beta}^\top(t)
\mathcal{U}_A\boldsymbol{h}\mathcal{U}_A^\dagger
=\boldsymbol{\beta}^\top (t) M_a\boldsymbol{h},$
where
\begin{equation}\label{Eq:MaDefinition}
    M_a=M_1M_2\dots M_n.
\end{equation}
By factorizing $\boldsymbol{h}$ and transposing we find
that
\begin{equation}
M_a^\top\boldsymbol{\beta}(t)=\boldsymbol{\beta}(t).\label{ma:eigval01}
\end{equation}
This means that $\boldsymbol{\beta}(t)$ is any eigenvector
of $M_a^\top$ with eigenvalue equal to 1, therefore,
in general
\begin{equation}
    \boldsymbol{\beta}(t)=\sum_{k=1}^m\gamma_k(t)\boldsymbol{\rho}_k(t),
    \label{ma:eigval02}
\end{equation}
where $\gamma_k(t)$ are coefficients to be determined
and $\boldsymbol{\rho}_k(t)$ are the eigenvectors of $M_a^\top$
whose eigenvalues are $1$. 
This equation directly provides a relation between
the components of $\boldsymbol{\beta}(t)$ and the $\boldsymbol{\alpha}(t)$
and reduces the search of parameters to $\gamma_1(t)$, $\dots$,
$\gamma_m(t)$ where $m<n$.

Summarizing, the method to determine $H_{\mathrm{e}}$ works as follows. 
1) Calculate the time-dependent $\boldsymbol{\alpha}(t)$
parameters by using Eq. (\ref{ode:01}) with
the initial condition $\boldsymbol{\alpha}(0)=0$.
2) Connect $\boldsymbol{\alpha}(t)$
and $\boldsymbol{\beta}(t)$ by means of
the solution of the ODE system (\ref{rel:01}) in the form
(\ref{rel:03}) and, if necessary,
use the eigenvalue one eigenvectors of $M_a^\top$ in
Eq. (\ref{ma:eigval02}) to simplify
the inverse relation (\ref{rel:03}).
3) Finally, $H_{\mathrm{e}}$
is obtained from (\ref{he:02}).

In what follows, we apply the method to three
well known problems: for the first one (Paul trap), only approximate 
solutions are known and the last two
of them (modulated optical lattice and
the Kapitza pendulum) have closed solutions. Here we find exact
solutions for the three of them.
As this method is rather systematic,
it can be put in the form of a symbolic computational algorithm  in Mathematica \cite{mathematica}.
The algorithms are provided in the supplemental material (SM) \cite{supmat}.

\noindent \emph{Example 1: Paul trap -}
Ion traps use time-dependent electric fields
in the radio frequency domain \cite{refId0,PhysRevX.4.031027}
to confine charged ions.
They are often studied through the Hamiltonian
of a particle of mass $m$ in a modulated harmonic potential
\begin{equation}
H=H_0+V(t)=\frac{1}{2m}p^2+\frac{m}{2}\left[\omega_1^2
+\omega_0^2\cos(\omega t)\right]x^2.\label{pt:01}
\end{equation}
The natural frequencies of the
constant and modulated potentials
are $\omega_1$ and $\omega_2$, respectively,
and $\omega$ is the radio angular frequency.
It can be easily shown that the operators
that constitute (\ref{pt:01}) form a Lie algebra.
The commutators of $h_1=x^2$, $h_2=p^2$ and $h_3=xp+px$
are $\left[x^2,p^2\right]=i\hbar 2\left(xp+px\right)=h_3$,
$\left[x^2,xp+px\right]=i\hbar 4x^2=h_1$,
$\left[p^2,xp+px\right]=-i\hbar 4p^2=h_2$.
Hence, its structure constants
are $c_{1,2,3}=-c_{2,1,3}=2$, $c_{1,3,1}=-c_{3,1,1}=4$
and $c_{2,3,1}=c_{3,2,1}=-4$.
This algebra corresponds to the
generators of the SU(2) group \cite{0305-4470-21-22-015}.

\begin{figure}
\includegraphics[width=0.30\textwidth, keepaspectratio=true]
{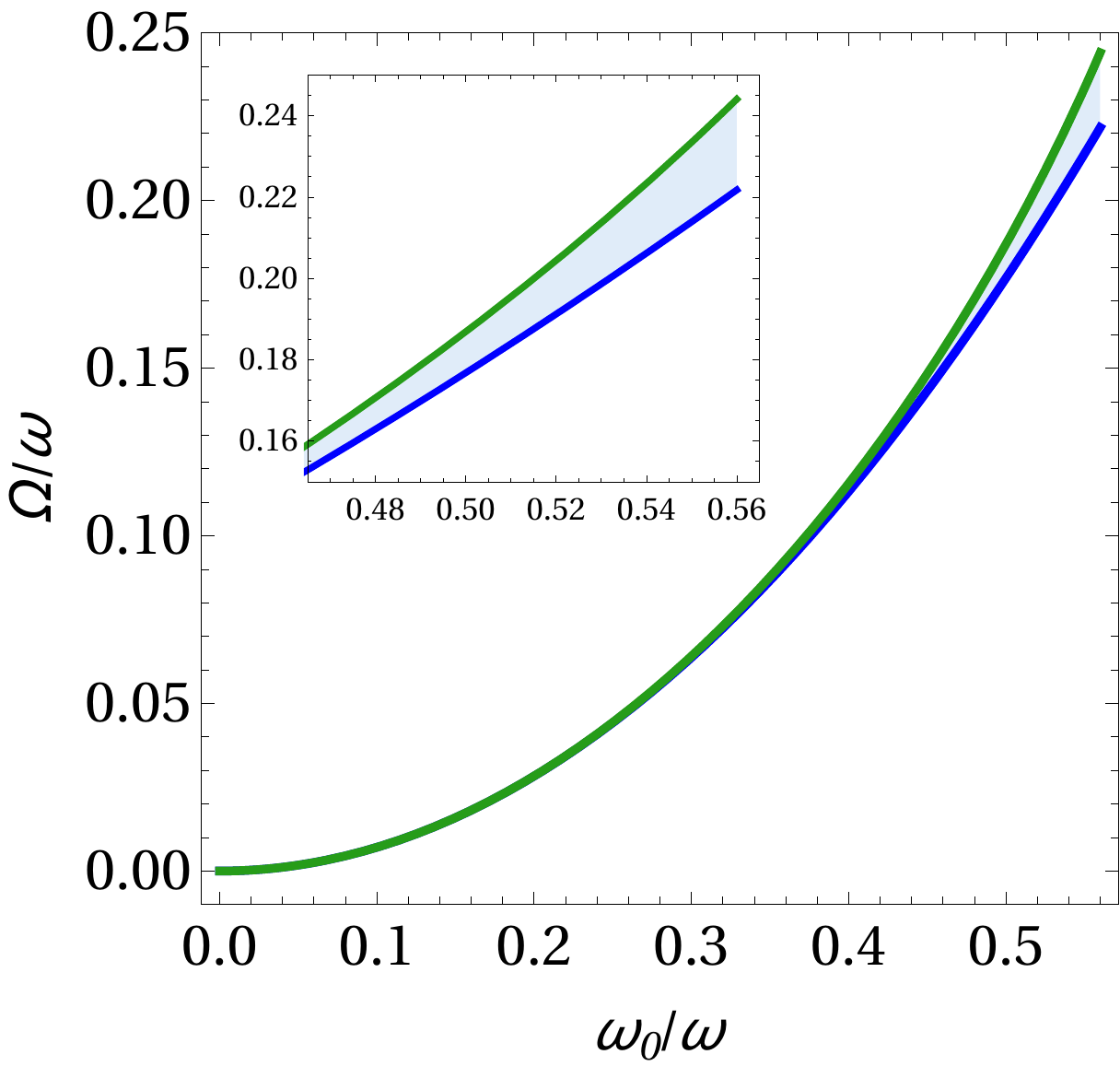}
\includegraphics[width=0.30\textwidth, keepaspectratio=true]
{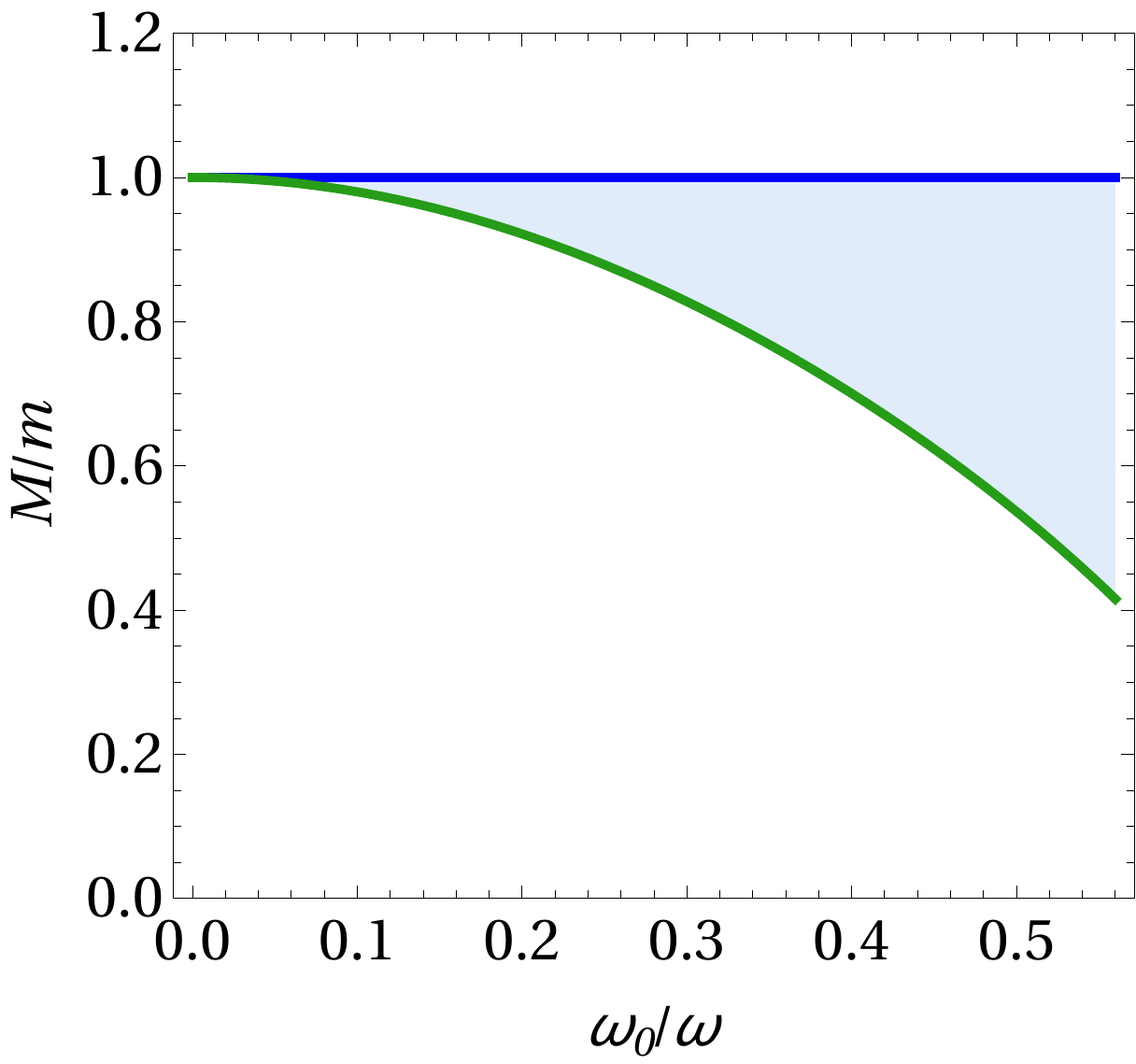}
\caption{Paul trap effective energy $\hbar\Omega$ (a) 
and effective mass $M$ (b) as function of
the drive's frequency $\omega_0/\omega$ obtained from Eq. (\ref{HePaul}). The green 
solid curves show the exact results and
the blue solid curves show the approximation
at first order ($H_0\ll V$). } 
\label{figure1}
\end{figure}

As shown in the SM, the solution resulting
from the ODE time-dependent transformation parameters is,
\begin{eqnarray} \label{Eq:PaulAlpha}
\alpha_1(t) &=& -\frac{m\omega^2}{8}
\frac{d}{dt}\ln C(a,q,\omega t/2),\\
\alpha_2(t) &=&  \frac{1}{2}\ln [C(a,q,\omega t/2)/C(a,q,0)],\\
\alpha_3(t) &=& \frac{C^2(a,q,0)}{2m}
\int_0^t\frac{ds}{C^2(a,q,\omega s/2)},
\end{eqnarray}
where $C(a,q,\omega t/2)$
is the even Mathieu function with $a=4\omega_1^2/\omega^2$ and $q=-2\omega_0^2/\omega^2$.
In order to obtain the $\boldsymbol{\beta}(t)$ we
derive the ODE system for $\lambda$ from (\ref{rel:01})
\begin{eqnarray}
\partial_{\lambda}\alpha_1(\lambda,t) &=&
  \beta_1(t) \mathrm{e}^{-4 \alpha_2(\lambda,t)}, \\
  \partial_{\lambda}\alpha_2(\lambda,t) &=&
  \beta_2(t)-2\beta_1(t) \alpha_3(\lambda,t), \\
\partial_{\lambda}\alpha_3(\lambda,t) &=&
  4 \beta_1(t) \alpha_3^2(\lambda,t)+\beta _3(t)\nonumber\\
  &&\,\,\,\,\,\,\,\,\,
  -4 \beta_2(t) \alpha_3(\lambda,t).\label{paultrap:lamb:01}
\end{eqnarray}
To avoid solving the whole system of differential
equations we may use the only eigenvalue one eigenvector of $M_a^{\top}$, given
in the SM. Therefore
\begin{equation}
\boldsymbol{\beta}(t) =\gamma_1(t) \left(
  \frac{\alpha_1(t)}{\alpha_3(t)},
  \frac{4 \alpha_1(t) \alpha_3(t)
  -\mathrm{e}^{-4 \alpha_2(t)}+1}{4 \alpha_3(t)},1\right),
  \label{paultrap:beta}
\end{equation}
where the explicit form of $\gamma_1(t)$ is given
in the SM.
Substituting the three components of $\boldsymbol{\beta}$
we finally obtain
the effective Hamiltonian
\begin{multline}
H_{\mathrm{e}}=\frac{\gamma_1(T)}{T}
    \bigg[
    p^2+
    \frac{\alpha_1(T)}{\alpha_3(T)}x^2\\
    +\frac{4\alpha_1(T) \alpha_3(T)
    -\mathrm{e}^{-4 \alpha _2(T)}+1}{4 \alpha_3(T)}
    (xp+px)
    \bigg].\label{paultrap:He}
\end{multline}
To first order in $H_0$ ($H_0\ll V$) the effective Hamiltonian is given
by $H_{\mathrm{e}}=p^2/2m+x^2m\omega_0^4/4\omega^2$ (SM)
in full consistency with
\cite{PhysRevX.4.031027}.
Even though the effective Hamiltonian in Eq. (\ref{paultrap:He})
is exact, it can be recast
in a more suitable form
as to allow the computation of the quasi-energies.
Applying the unitary transformation
$U\equiv U_1(\beta_2(t)/2\beta_3(t))=\exp(ix^2\beta_2(t)/2\beta_3(t)\hbar)$
the effective Hamiltonian is transformed into
\begin{equation}\label{HePaul}
    H^\prime_{\mathbf{e}}=UH_{\mathbf{e}}U^\dagger=
      \frac{\beta_3(T)}{T} p^2
      +\frac{1}{T}\left[\beta_1(T)-\frac{\beta_2^2(T)}{\beta_3(T)}\right]x^2,
\end{equation}
where $\beta_1(T)$, $\beta_2(T)$ and $\beta_3(T)$ are readily
obtained from (\ref{paultrap:beta}).
Figures \ref{figure1} (a) and (b) exhibit the behaviour
of the effective energy
$\hbar\Omega/\hbar\omega=\sqrt{\beta_1\beta_3-\beta_2^2}/\pi$
and mass $M/m=\pi/m\omega\beta_3$ as functions
of the drive's frequency $\omega_0/\omega$.
The green solid lines show the exact calculations and
the blue ones show the results corresponding to
the approximation $H_0\ll V$, $M/m=1$ and
$\hbar \Omega/\hbar \omega=\omega_0^2/\sqrt{2}\omega^2$. We observe that
for small values of $\omega$ the exact and approximate
solutions of $\Omega/\hbar \omega$  slightly diverge.
The exact effective mass, on the other hand, is rather different
from the approximated one, even for small values of $\omega/\omega_0$.

\noindent \emph{Example 2: Modulated optical lattice-}
The second-quantized tight-binding Hamiltonian of
the modulated optical lattice
\cite{PhysRevLett.99.220403,PhysRevX.4.031027}
is given by
\begin{equation}
H=H_0+\omega\kappa \cos\left(\omega t\right) V,
\end{equation}
where $\kappa$ is a constant parameter,
$H_0=J\sum_j(a_{j+1}^\dagger a_j+a_{j}^\dagger a_{j+1})$
is the nearest-neighbor hopping term and
$V=\sum_jja_j^\dagger a_j$ is
the lattice potential.
The operators $a_j^\dagger$ and  $a_j^\dagger$
are standard boson creation and annihilation operators at cite $j$.
Following the procedure described above
the effective Hamiltonian is
found to be
\begin{multline*}
H_{\mathrm{e}}=\Big(
   \beta_1(T)h_1
  +\beta_2(T)h_2
  +\beta_3(T)h_3\Big)/T
=J_0(\kappa)H_0,
\end{multline*}
where $\beta_1(T) =$ $\beta_2(T)=0,$ and $\beta_3(T)=TJ_0(\kappa)$.
A detailed calculation of these parameters can
be found in the SM.
$H_{\mathrm{e}}$ is the same as the exact solution given in Ref. \cite{PhysRevX.4.031027}.

\noindent \emph{Example 3: Kapitza pendulum-}
Here we examine the Hamiltonian of a
harmonic oscillator subject to a time-dependent force
\cite{PhysRevA.68.013820}
\begin{equation}
    H=\frac{p^2}{2m}+\frac{1}{2}m\omega_0^2x^2+x F\cos(\omega t).
    \label{kapitza:ham01}
\end{equation}
In principle, the three elements in this Hamiltonian
can be identified as part of the algebra formed by the operator set
$ h_1 = 1$,
$ h_2 = x$,
$ h_3 = p$ ,
$ h_4 = x^2$,
$ h_5 = xp +px$,
$ h_6 = p^2$
However, calculations are sizeabley simplified
by choosing instead
$ h_1 = 1$,
$ h_2 = x$,
$ h_3 = p$ ,
$ h_4 = m^2\omega_0^2x^2 +p^2$.
The corresponding non-vanishing structure constants are
$c_{2,3,1}= -c_{3,2,1}=1$, $c_{4,2,3}= -c_{2,4,3}=-2$,
$c_{4,3,2}=-c_{3,4,2}=2m^2\omega_0^2$. By following the method, as detailed in the SM, the effective Hamiltonian is
\begin{multline}
H_{\mathrm{e}}=\Big(
  \beta_1(T) +\beta_2(T) x
  +\beta_3(T) p
  \\
  +\beta_4(T)\left[ p^2+ (m \omega_0)^2 x^2\right] \Big)/T,  
\label{kapitza:He}
\end{multline}
where $\beta_1(T)$, $\beta_2(T)$,
$\beta_3(T)$ and $\beta_4(T)$ are explicitly
given in the SM.
This Hamiltonian can be rewritten in a more familiar
form by eliminating the terms proportional to $x$ and $p$
via the unitary transformation
$U\equiv U_2(\beta_3/2\beta_4)
U_3(-\beta_2/2(m\omega_0)^2\beta_4)
=\exp(i\hbar x \beta_3/2\beta_4)
\exp(-i\hbar p \beta_2/2(m\omega_0)^2\beta_4)$.
The transformed effective Hamiltonian takes the form
\begin{equation}
H^\prime_e =UH_{\mathrm{e}}U^\dagger
  = \frac{p^2}{2m}+ \frac{1}{2} m \omega_0^2 x^2 
    + \frac{  F^2}{4 m \left(\omega^2 -\omega _0^2\right)}.
    \label{kapitza:hame01}
\end{equation}
Though this effective Hamiltonian has not been
determined explicitly before, (\ref{kapitza:hame01}) is consistent with its
very well known  quasienergies
\cite{PhysRevA.68.013820}.

In conclusion, we have presented a general method to find the time evolution operator and the effective Hamiltonian for time-driven systems using an algebraic approach. Then we reproduced the solutions for known exact solvable models, while we solved the Paul trap model.

This work was supported by DCB UAM-A grant numbers
2232214 and 2232215, and UNAM DGAPA PAPIIT IN102717. J.C.S.S. has a scholarship from Becas de Posgrado UAM number 2151800745.


%

\end{document}